## 1. TITLE

**Contrast detection is enhanced by deterministic, high-frequency transcranial alternating current stimulation with triangle and sine waveform.**

## 2. ABBREVIATED TITLE

**tACS modulates visual processing of V1.**

## 3. AUTHORS AND AFFILIATIONS

Weronika Potok[1,2*], Onno van der Groen[4], Sahana Sivachelvam[1], Marc Bächinger[1,2], Flavio Fröhlich[5-10], Laszlo B. Kish[11], Nicole Wenderoth[1,2,3*]

[1] Neural Control of Movement Lab, Department of Health Sciences and Technology, 8093, Zurich, Switzerland
[2] Neuroscience Center Zurich (ZNZ), University of Zurich, Federal Institute of Technology Zurich, University and Balgrist Hospital Zurich, 8057, Zurich, Switzerland
[3] Future Health Technologies, Singapore-ETH Centre, Campus for Research Excellence And Technological Enterprise (CREATE), 138602, Singapore.
[4] Neurorehabilitation and Robotics Laboratory, School of Medical and Health Sciences, Edith Cowan University, 6027, Joondalup, Australia
[5] Department of Psychiatry, University of North Carolina at Chapel Hill, Chapel Hill, NC, USA
[6] Carolina Center for Neurostimulation, University of North Carolina at Chapel Hill, Chapel Hill, NC, USA
[7] Department of Neurology, University of North Carolina at Chapel Hill, Chapel Hill, NC, USA
[8] Department of Cell Biology and Physiology, University of North Carolina at Chapel Hill, Chapel Hill, NC, USA
[9] Department of Biomedical Engineering, University of North Carolina at Chapel Hill, Chapel Hill, NC, USA
[10] Neuroscience Center, University of North Carolina at Chapel Hill, Chapel Hill, NC, USA
[11] Department of Electrical & Computer Engineering, Texas A&M University, TAMUS 3128, College Station, TX 77843-3128, USA

## 4. AUTHORS' CONTRIBUTIONS

Weronika Potok and Onno van der Groen have made an equal contribution.
WP, OvdG, SS, MB, LBK, and NW designed the study
WP and SS collected the data
WP and SS analyzed the data
WP, OvdG, FF, LBK, and NW wrote the paper

## 5. * CORRESPONDING AUTHORS

Weronika Potok and Prof. Nicole Wenderoth

Neural Control of Movement Laboratory
Department of Health Sciences and Technology
ETH Zurich, Switzerland
Auguste-Piccard-Hof 1, 8093 Zurich
Phone: +41 44 633 29 37
E-mail: weronika.potok@hest.ethz.ch, nicole.wenderoth@hest.ethz.ch

**6. NUMBER OF FIGURES:** 7
**7. NUMBER OF TABLES:** 0
**8. NUMBER OF MULTIMEDIA:** 0
**9. NUMBER OF WORDS FOR ABSTRACT:** 236 w.
**10. NUMBER OF WORDS FOR NEW & NOTEWORTHY:** 54 w.
**11. NUMBER OF WORDS FOR INTRODUCTION:** 2501 w.
**12. NUMBER OF WORDS FOR DISCUSSION:** 2107 w.

**13. ACKNOWLEDGEMENTS:** We thank all the participants for their time and effort.

**14. DECLARATIONS OF INTEREST**: Authors report no conflict of interest

## 15. FUNDING SOURCES

This work was supported by the Swiss National Science Foundation (320030_175616) and by the National Research Foundation, Prime Minister's Office, Singapore under its Campus for Research Excellence and Technological Enterprise (CREATE) programme (FHT).

**Abstract**

Stochastic Resonance (SR) describes a phenomenon where an additive noise (stochastic carrier-wave) enhances the signal transmission in a nonlinear system. In the nervous system, nonlinear properties are present from the level of single ion channels all the way to perception and appear to support the emergence of SR. For example, SR has been repeatedly demonstrated for visual detection tasks, also by adding noise directly to cortical areas via transcranial random noise stimulation (tRNS). When dealing with nonlinear physical systems, it has been suggested that resonance can be induced not only by adding stochastic signals (i.e., noise) but also by adding a large class of signals that are not stochastic in nature which cause "deterministic amplitude resonance" (DAR). Here we mathematically show that high-frequency, deterministic, periodic signals can yield resonance-like effects with linear transfer and infinite signal-to-noise ratio at the output. We tested this prediction empirically and investigated whether non-random, high-frequency, transcranial alternating current stimulation applied to visual cortex could induce resonance-like effects and enhance performance of a visual detection task. We demonstrated in 28 participants that applying 80 Hz triangular-waves or sine-waves with tACS reduced visual contrast detection threshold for optimal brain stimulation intensities. The influence of tACS on contrast sensitivity was equally effective to tRNS-induced modulation, demonstrating that both tACS and tRNS can reduce contrast detection thresholds. Our findings suggest that a resonance-like mechanism can also emerge when deterministic electrical waveforms are applied via tACS.

**Keywords**

**New & Noteworthy**

Our findings extend our understanding of neuromodulation induced by noninvasive electrical stimulation. We provide first evidence showing acute online benefits of $tACS_{triangle}$ and $tACS_{sine}$ targeting the primary visual cortex (V1) on visual contrast detection in accordance with the resonance-like phenomenon. The 'deterministic' tACS and 'stochastic' hf-tRNS are equally effective in enhancing visual contrast detection.

# 1. Introduction

## 1.1 On stochastic resonance

Stochastic resonance (SR) was discovered in the context of the hysteresis features of climate (ice age) (1–3). Since then it has been generalized and studied in a variety of naturally occurring processes including biological systems (4, 5). Demonstrations of SR in the nervous system were carried out on crayfish mechanoreceptors (6), neurons in crickets (7), mice (8, 9), rats (10–12), cats (13), and humans (see *1.3 Stochastic resonance effects on neural processing* below) with studies consistently reporting enhanced system performance. Signal enhancement is described in a vast body of literature as the basic property of resonance mechanisms (14). Here we survey a few basic features of SR that are directly relevant for our paper. In general, the quality of signal transfer through a system is characterized by the following parameters at the output: amplification (or the signal strength), linearity, signal-to-noise ratio, and the phase shift.

SR is a phenomenon where the transfer of a periodic or aperiodic signal in a nonlinear system is optimized by an additive -typically Gaussian- noise (15). Note that originally, when SR was studied in binary systems, it represented a frequency-resonance, that is, matching the period time of the periodic signal with the mean residence time in the potential wells of the binary system driven by a stochastic carrier-wave (noise). Later the argument behind the name SR was modified to amplitude-resonance. Today, "resonance" means an optimal root-mean-square (RMS) amplitude value of the noise, i.e., amplitude-resonance at the carrier-wave RMS amplitude level for the best signal transmission.

In the initial phase of SR research, the nonlinear systems were bistable (1). At a later stage it was discovered that monostable systems (including neurons) also offer SR (16). Moreover, it was realized that the memory/hysteresis effects of the bistable systems actually cause a stochastic phase shift (phase noise) that negatively impacts the quality of the transferred signal (17). Due to this fact, the best stochastic resonators are the memory-free Threshold Elements (TE), such as the Level Crossing Detector (LCD) (18) and the Comparator (19). The LCD device (the simplest model of a neuron) produces a short, uniform spike whenever its input voltage amplitude is crossing a given threshold level in a chosen, typically positive direction. On the other hand, the Comparator has a steady binary output where the actual value is dictated by the situation of the input voltage amplitude compared to a given threshold level: for example, in the sub-threshold case the output is "high" while in the supra-threshold case, it is "low".

At the output of a stochastic resonator, the signal strength (SS), the signal-to-noise-ratio (SNR), the information entropy and the Shannon information channel capacity show maxima versus the intensity of the additive input noise. However, these maxima are typically located at different noise intensities. Exceptions are the SNR and information entropy which are interrelated by a monotonic function; thus they have the same location of their maxima, see the arguments relevant for neural spike trains (20). On the other hand, the information channel capacity of SR in an LCD and in neural spike trains has the bandwidth as an extra variable controlled by the input (the higher the input noise the higher the bandwidth); thus the different location of its maximum is at higher input noise than for the maximum of the SNR (21).

It is important to note that, *in the linear response limit*, that is, when the input signal is much smaller than the RMS amplitude of the additive carrier-wave (Gaussian noise), the SNR at the output is always less than at the input (see the mathematical proof in (15)). Consequently, the information content at the output is always less than at the input. On the contrary, in the nonlinear response limit, the SNR at the output can be enhanced by several orders of magnitude compared to its input value provided the signal has a small duty cycle, such as neural spikes do (17, 22). Yet, due to the unavoidable noise at the output, which is the unavoidable impact of the stochastic carrier-wave (noise), the information at the output (signal plus noise) is always less than in the *original* input signal *without the added carrier-wave* (*noise*).

Therefore, if a proper additive, high-frequency, periodic time function could be used as carrier-wave in a stochastic resonator instead of a Gaussian noise, the fidelity and the information content of the input signal could be preserved while it is passing through the nonlinear device, as we will show below. However, even in this case there is an optimal (range) for the mean-square amplitude of the carrier-wave. Thus, we call this deterministic phenomenon "*deterministic amplitude resonance"*, (DAR), which is also an amplitude-resonance where a *deterministic* (instead of stochastic/noise) carrier-wave with sufficiently large amplitude produces the optimal signal transfer via the system.

## 1.2 Deterministic amplitude resonance (DAR) with high-frequency periodic carrier-waves

First Landa and McClintock (23) realized that SR like phenomena could occur with high-frequency sinusoidal signals instead of noise. They successfully demonstrated their idea by computer simulations of a binary (double-well potential) SR system. Recently, Mori, et al (24) used high-frequency, noise-free, periodic neural spikes for excitation in a neural computer model to show that SR like features on the mutual information can be achieved by tuning the frequency of these periodic excitation in the 80-120 Hz range.

Below, we show that high-frequency triangle waves can offer a noise-free signal transfer which can be exactly linear at certain conditions. Sinusoidal waves are also discussed briefly.

### *1.2.1 The case of triangle (or sawtooth) carrier-waves, instead of noise*

Earlier, in a public debate about the future of SR, one of us proposed a noise-free method by utilizing high-frequency triangle waves to improve signal transmission through threshold devices (25) and to reach exactly linear transfer and infinite SNR at the output. Here we summarize those arguments.

**Figure 1** shows an example of stochastic resonator hardware with an additive triangle wave, as the carrier-wave, instead of noise. The same argumentation works for sawtooth wave, too. Note: the original threshold-based stochastic resonators (17, 18) contain the same hardware elements where Gaussian random noise is used instead of the triangle wave. Due to the binary nature of the visual detection experiments described in this paper our focus is on sub-threshold binary (square-wave) signals with some additional comments about the case of analog signals.

#### FIGURE 1 ###

The TE is either an LCD or a Comparator. Suppose that the stable output of the LCD is zero and it produces a short uniform positive spike with height $U_{\text{LCD}}$ and duration $\tau$ whenever the input level crosses the Threshold in upward direction. The Comparator's output stays at a fixed positive value whenever the input level is greater than the threshold and stays at a lower value (zero or negative) otherwise. Suppose when the input level is greater than the Threshold, $U_{\text{th}}$, the Comparator output voltage $U_{\text{c}} = U_{\text{H}}$ and otherwise it is 0. The Low-pass Filter creates a short-time moving-average in order to smooth out the high-frequency components (frequency components due to switching triggered by the carrier wave) and it keeps only the low-frequency part which is the bandwidth of the signal. The parameters, such as the frequency $f_{\text{s}}$ of the signal, the frequency $f_{\text{t}}$ of the triangle wave and the cut-off frequency $f_{\text{c}}$ of the Low-pass Filter should satisfy

$$f_{\text{s}} \ll f_{\text{c}} \ll f_{\text{t}} < \frac{1}{\tau} \tag{1}$$

in order to transfer the signal with the highest fidelity.

The upper part of **Figure 1** shows the situation without carrier wave: the sub-threshold binary signal is unable to trigger the TE thus the output signal is steadily zero. The lower part of **Figure 1** shows the situations where an additive, triangle wave assists the signal to reach the threshold thus it carries the binary signal over the TE resulting in a nonzero output signal. The triangular wave will have to be of a high enough frequency for two main reasons i) because of the Nyquist sampling theorem, the sampling frequency needs to be at least twice as large than the highest frequency component of the signal, ii) for the low-pass filter to be able to smooth out the carrier signal (which is a trash), the carrier signal frequency must be much larger than the reciprocal of the time duration of the binary signal.

i) The case of Level Crossing Detector (LCD)

If a constant input signal plus triangle wave can cross the threshold, the LCD produces a periodic spike sequence with the frequency of the triangle wave. In this situation, the time average of this sequence is $f_t \tau U_{\mathrm{LCD}}$ therefore, for the binary input signal, the output of the LPF will be binary with amplitude values:

$$U_{\mathrm{LPF}}(t) = f_t \tau U_{\mathrm{LCD}} \quad \text{or} \quad 0 \tag{2}$$

Thus, the binary input signal is restored at the output of the LPF without any stochasticity (noise) in it. The only deviation from the input signal is a potentially different amplitude (non-zero amplification) and some softening of the edges dues to the LPF depending on how well Relation 1 is satisfied.

In conclusion, with an LCD as TE, regarding the amplitude resonance versus the carrier wave amplitude $U_t$, there are three different input amplitude ranges:

(a) $U_s + U_t < U_{\mathrm{th}}$    then there is no output signal

(b) $U_{\mathrm{th}} < U_s + U_t$ , $U_s < U_{\mathrm{th}}$ , $U_t < U_{\mathrm{th}}$   then the binary signal is restored at the output

(c) $U_{th} < U_t$    then the output is steadily at the high level $U_{LPF}(t) = f_t \tau U_{LCD}$

Therefore, the binary signal can propagate to the output only in the (b) situation when it does that without any noise contribution at the output (the SNR is infinite).

ii) The case of Comparator

Note, this system is very different from "Stocks's suprathreshold SR" (19), where a large number of independent comparators with independent noises are used with a common signal and an adder to reach a finite SNR. For the sake of simplicity, but without limiting the generality of the argumentation, suppose that the binary signal, $U_s(t)$, values are 0 and $U_{\check{s}}$, where $U_s \leq U_{th}$, and the maximum amplitude of the triangle signal, $U_t(t)$   is $U_t$ and its minimum value is $_0$. In conclusion:

$$\text{when } U_s(t) + U_t(t) > U_{th}   ,   U_c = U_H \text{ otherwise   } U_c = 0 \qquad (3)$$

i.e., the comparator's output voltage has only 2 possible steady states: 0 and $U_H$. When the input voltage is above the threshold voltage $U_{th}$ the output is $U_H$, otherwise it is 0.

To evaluate the average output voltage of the comparator the first question is the fraction of time that the input spends over the threshold, see **Figure 2**.

### FIGURE 2 ###

This time $T_H$ within a period of the triangle wave is the period duration $1/f_t$ minus twice the time $t_r$ spent for rising from the minimum to the threshold:

$$T_H = \frac{1}{f_t} - 2 t_r = \frac{1}{f_t} - 2 \frac{U_{th} - U_s}{2 U_t f_t} = \frac{U_t - U_{th} - U_s}{U_t f_t}, \qquad (4)$$

where we used that the slope $s$ of the triangle signal with peak-to-peak amplitude is

$$s = 2 f_t U_t \quad , \tag{5}$$

assumed that that the signal amplitude $U_s$ is present at the input and assumed condition (3) that the signal alone is subthreshold, but the sum of the signal and the triangle wave is suprathreshold:

$$\tag{6}$$

From (3) and (4), the smoothed value of the output voltage $U_{LPF}(t)$ of the LPF when the input signal amplitude is :

$$U_{LPF} = \langle U_{LPF}(t) \rangle = U_H \frac{T_H}{1/f_t} = U_H \frac{U_t - U_{th} + U_s}{U_t} = U_H \frac{U_t - U_{th}}{U_t} + \frac{U_H}{U_t} U_s \quad , \tag{7}$$

where denotes short-range averaged (smoothed) value discussed above.

The last term in the right side of Equation (7) demonstrates that the signal amplitude transfers linearly through the system. Therefore, this version of our device is working distortion-free also for analog signals, not only for the present digital signal assumption.

Thus, this device is not only noise-free but also ideally linear for subthreshold signals satisfying condition (6), even though exact linearity is not an important feature during the experimental study in the present paper.

In conclusion, with a comparator as TE, regarding the amplitude resonance versus the carrier wave amplitude , there are two different input amplitude ranges:

(a) $U_s + U_t < U_{th}$ then there is no output signal

(b) $U_{th} < U_s + U_t$ , $U_s < U_{th}$, then the binary signal is restored at the output and its amplitude scales inversely with the amplitude $U_t$ of the carrier wave. The maximal amplitude is at .

Therefore, the binary signal can propagate to the output only in the (b) situation when it does that without any noise (the SNR is infinite) and it has a linear transfer for analog signals. Of note, for a large $U_t$ (trianguler waveform) the binary signal is lost as the output would be a constant signal.

### 1.2.2 The case of sinusoidal carrier waves, instead of triangle waves

The above argumentations qualitatively work also for sinusoidal carrier-waves except that the linearity of the transfer is lost. The triangle carrier-wave has a Fourier series that has only odd harmonics, where the *n*-th harmonic amplitudes scale with $1/n^2$, that is, the strongest harmonic (the 3-rd) is 9 times weaker, and the next strongest harmonic (the 5-th) is 25 times less than the base harmonic. The qualitative difference is that the absolute value of the slope of sinusoidal carrier-wave is reduced when approaching its peak level and it is zero at the peak. The constant slope of the triangle wave is essential for the exactly linear transfer, see the mathematical proof above.

In conclusion, when sinusoidal carrier-wave is used instead of a triangle (or sawtooth) wave, the same qualitative features remain, including the zero-noise contribution at the output (infinite SNR). The exception is the linearity of transfer of analog signals via the comparator which is lost at sinusoidal carrier-wave.

## 1.3 Stochastic resonance effects on neural processing

In neural systems, it has been demonstrated that responses to externally applied stimuli were maximally enhanced when an optimal level of electrical random noise stimulation was applied. These effects were linked specifically to the opening of voltage gated sodium ($Na^+$) channels in response electrical stimulation, causing a sodium influx, which in turn causes a local depolarization of the cell membrane (9, 12, 26).

In humans, early SR effects have been mainly demonstrated via behavioral signal detection tasks whereby noise was added to the periphery. For example, the detection of low-contrast visual stimuli was significantly enhanced when the stimuli were superimposed with visual noise (27)

Recently, similar enhancements of visual perception have been reported when noise was directly added to the cerebral cortex by the means of transcranial random noise stimulation (tRNS) in studies investigating its acute effects on visual processing (26, 28–33). According to the SR theory, while the optimal level of tRNS benefits performance, excessive noise is detrimental for signal processing (28, 29, 33), resulting in an inverted U-shape relationship between noise benefits and noise intensity. In consistence with SR, tRNS was shown to be particularly beneficial for visual detection performance when the visual stimuli were presented with near-threshold intensity (28, 29, 34).

However, based on the theoretical consideration described above, a resonance-like phenomenon can be observed with deterministic stimulations. Here we test this prediction empirically and investigate if the response of visual cortex to around-threshold contrast stimuli could also be enhanced via a high-frequency deterministic signal. We tested if triangle or sine waves can modulate signal processing in a resonance-like manner by delivering tACS with triangle waveform (tACS$_{triangle}$) or sine waveform (tACS$_{sine}$) targeting the primary visual cortex (V1) of participants performing a visual contrast sensitivity task and measured their visual detection threshold. We hypothesized that resonance-like DAR effects would be reflected in the beneficial influence of high-frequency stimulation on signal processing via signal enhancement.

## 2. Materials and methods

## 2.1 Participants

Individuals with normal or corrected-to-normal vision and without identified contraindications for participation according to established brain stimulation exclusion criteria (35, 36) were recruited in the study. All study participants provided written informed consent before the beginning of each experimental session. Upon study conclusion participants were debriefed and financially compensated for their time and effort. All research procedures were approved by the Cantonal Ethics Committee Zurich (BASEC Nr. 2018-01078) and were performed in accordance with the Helsinki Declaration of the World Medical Association (2013 WMA Declaration of Helsinki) and guidelines for non-invasive brain stimulation research through the COVID-19 pandemic (37).

The required sample size was estimated using an a priori power analysis (G*Power version 3.1; (38)) based on the effect of maximum contrast sensitivity improvement with tRNS shown by Potok et al. (39) ($\eta_p^2$ = 0.165, Effect size f = 0.445). It revealed that 28 participants should be included in an experiment to detect an effect with repeated measures analysis of variance (rmANOVA, 4 levels of stimulation condition), alpha = 0.05, and 90% power. We included 31 participants in experiment 1 (tACS$_{triangle}$) and 32 participants in experiment 2 (tACS$_{sine}$) to account for potential dropouts. Visual contrast detection is potentially prone to floor effects if the contrast detected at baseline approaches the technical limits of the setup. We decided to exclude participants that are exceptionally good in the visual task and present visual contract threshold below 0.1 (Michelson contrast, see *Visual stimuli*) in the baseline, no tACS condition (for visual contrast intensity range of minimum 0 and maximum 1) in the no tACS baseline condition. For outliers' removal we used standardized interquartile range (IQR) exclusion criteria (values below Q1-1.5IQR or above Q3+1.5IQR, where Q1 and Q3 are equal to the first and third quartiles, respectively) to avoid accidental results, unlikely driven by tES, e.g., due to participants responding without paying attention to the task.

From the initially recruited sample, we excluded 7 individuals. In tACS_triangle experiment 1: 1 participant revealed exceptional contrast threshold modulation (>Q3+1.5IQR), 1 participant had a contrast threshold below 0.1 in the baseline condition (also >Q3+1.5IQR), 1 participant stopped the session because of unpleasant skin sensations. In tACS_sine experiment 2: 1 participant revealed exceptional contrast threshold modulation (>Q3+1.5IQR), 1 participant stopped the session because of unpleasant skin sensations, 2 participants reported frequent (75% accuracy) phosphenes sensation due to stimulation (see *tACS characteristics*).

The final sample consisted of 28 healthy volunteers (16 females, 12 males; 26.9 ± 4.7, age range: 21-39) in tACS_triangle experiment 1, and 28 healthy volunteers (20 females, 8 males; 26.4 ± 4.4, age range: 20-39) in tACS_sine experiment 2. We did not collect information about the race of participants. Twenty of these participants completed both experimental sessions. For participants who took part in both experiments, 15 participants started with tACS_triangle and 5 with tACS_sine. The experimental sessions took place on different days with 2.6 ± 1.2 months on average apart. Delays were caused by COVID-19 pandemic (37).

## 2.2 General Study design

To evaluate the influence of tACS on visual contrast detection, we performed two experiments in which we delivered either tACS_triangle, or tACS_sine targeting V1, during visual task performance (see **Figure 3A**). In each experiment, three tACS intensities and a control no tACS condition were interleaved in a random order. Our main outcome parameter in all experiments was a threshold of visual contrast detection (VCT) that was determined for each of the different tACS conditions (39). The experimental procedure to estimate VCT followed a previously used protocol to assess the influence of tRNS on contrast sensitivity (39). In brief, VCT was estimated twice independently, in two separate blocks within each session (see **Figure 3D**). We determined the individual's optimal tACS intensity (defined as the intensity causing the lowest VCT, i.e., biggest improvement in contrast sensitivity) for each participant in the 1st block of experiment 1 (ind-tACS_triangle) and experiment 2 (ind-tACS_sine) and retested their effects within the same experimental session on VCT data acquired in the 2<sup>nd</sup> block.

### FIGURE 3 ###

### 2.2.1 Experimental setup and visual stimuli

The experiments took place in a dark and quiet room, ensuring similar lighting conditions for all participants. Participants sat comfortably, 0.85m away from a screen, with their head supported by a chinrest. Visual stimuli were generated with Matlab (Matlab 2020a, MathWorks, Inc., Natick, USA) using the Psychophysics Toolbox extension that defines the stimulus intensity with Michelson contrast (40–42) and displayed on a CRT computer screen (Sony CPD-G420). The screen was characterized by a resolution of 1280 x 1024 pixels, refresh rate of 85Hz, linearized contrast, and a luminance of 35 cd/m$^2$ (measured with J17 LumaColor Photometer, Tektronix$^{TM}$). The target visual stimuli were presented in the form of the Gabor patch – a pattern of sinusoidal luminance grating displayed within a Gaussian envelope (full width at half maximum of 2.8 cm, i.e., 1° 53' visual angle, with 7.3 cm, i.e., 4° 55' presentation radius from the fixation cross, staying within the central vision, i.e., <8° radius; (43, 44)). The Gabor patch pattern consisted of 16 cycles with one cycle made up of one white and one black bar (grating spatial frequency of 8 c/deg). Stimuli were oriented at 45° tilted to the left from the vertical axis (see **Figure 3A**), since it was shown that tRNS enhances detection of low contrast Gabor patches especially for non-vertical stimuli of high spatial frequency (31).

### 2.2.2 Four-alternative forced choice visual detection task

In both experiments, participants performed a visual four-alternative forced choice (4-AFC) visual task, designed to assess an individual VCT, separately for each stimulation condition. A 4-AFC protocol was shown to be more efficient for threshold estimation than commonly used 2-AFC (45). Participants were instructed to fixate their gaze on a cross in the center of the screen. In the middle of each 2.04s trial, a Gabor patch was randomly presented for 40ms in one of the 8 locations (see **Figure 3A**). A stimulus appeared in each location for the same number of times (20) within each experimental block in pseudo-randomized order to avoid a spatial detection bias. The possible locations were set on noncardinal axes, as the detection performance for stimuli presented in this way is less affected (i.e. less variable) than when stimuli are positioned on the cardinal axes (46). Each trial was followed by 1s presentation of only fixation cross after which the 'response screen' appeared. Participants' task was to decide in which quadrant of the screen the visual stimulus appeared and indicate its location on a keyboard (see **Figure 3A**). The timing of the response period was self-paced and not limited. Participants completed a short training session (10 trials), with the stimulus presented always at high contrast (0.5; for visual contrast intensity range of minimum 0 and maximum 1), in order to ensure that they understand the task (see **Figure 3D**).

VCT was estimated using the QUEST staircase procedure (47), implemented in the Psychophysics Toolbox in Matlab (40–42), which is a method used in psychophysical research to estimate threshold of a psychometric function (47). The thresholding procedure starts with a presentation of the visual stimulus displayed with 0.5 contrast intensity (Michelson contrast, for visual contrast intensity ranging 0-1; note that the stimuli were displayed for just 40ms). When participants answer correctly, QUEST lowers the presented contrast intensity. Consequently, when participants answer incorrectly QUEST increases the presented contrast. The estimated contrast intensity for the next stimulus presentation is based on a maximum-likelihood-based estimate of the underlying psychometric function. Characteristics of the stimuli on each trial are determined by the input stimuli and respective system responses that occurred in the previous sequence of trials (48). The estimated stimulus contrast is adjusted to yield 50% detection accuracy (i.e., detection threshold criterion, see **Figure 3C**). For a 4-AFC task 25% accuracy corresponds to a chance level. The remaining parameters used in the QUEST staircase procedure where set as follows:

steepness of the psychometric function, beta = 3; fraction of trials on which the observer presses blindly, delta = 0.01; chance level of response, gamma = 0.25; step size of internal table grain = 0.001; intensity difference between the largest and smallest stimulus intensity, range = 1. VCT was assessed across 40 trials per stimulation condition. Four different conditions were randomly interleaved within each of 2 experimental blocks (40 trials x 4 conditions x 2 blocks; total number of 320 trials per experimental session, **Figure 3D**).

### 2.2.3 tACS characteristics

In stimulation trials, tACS (80Hz) with symmetrical triangle- (tACS$_{triangle}$) or sinewave (tACS$_{sine}$), with no offset was delivered. Stimulation started 20ms after trial onset and was maintained for 2s (see **Figure 3A**). Subsequently a screen with only fixation cross was displayed for 1 s, followed by the self-paced response time. tACS waveforms were created within Matlab (Matlab 2020a, MathWorks, Inc., Natick, USA) and sent to a battery-driven electrical stimulator (DC-Stimulator PLUS, NeuroConn GmbH, Ilmenau, Germany), operated in REMOTE mode, via a National Instruments I/O device USB-6343 X series, National Instruments, USA). The active tACS conditions and no tACS control condition were interleaved and presented in random order. Timing of the stimuli presentation, remote control of the tACS stimulator, and behavioral data recording were synchronized via Matlab (Matlab 2020a, MathWorks, Inc., Natick, USA) installed on a PC (HP EliteDesk 800 G1) running Windows (Windows 7, Microsoft, USA) as an operating system.

In both experiments tACS (80Hz) stimulation (tACS$_{triangle}$ in experiment 1 or tACS$_{sine}$ in experiment 2) was delivered with 0.75mA, 1mA, and 1.5mA amplitude (peak-to-baseline), resulting in maximum current density of 60 $\frac{\mu A}{cm2}$, which is below the safety limits of 167 $\frac{\mu A}{cm2}$ for transcranial electrical stimulation (49). These intensities were selected based on previous studies investigating effects of tRNS on contrast sensitivity (28, 39).

Prior to electrode placement, an anesthetic cream (Emla® 5%, Aspen Pharma Schweiz GmbH, Baar, Switzerland) was applied to the intended electrodes position on the scalp to numb potential tACS-induced cutaneous sensations and diminish transcutaneous effects of stimulation. To ensure that the cream got properly absorbed, it was left on the scalp for 20 min (50, 51) during which participants completed task training (see *Four-alternative forced choice visual detection task* and **Figure 3D**).

To target V1 we used an electrode montage that was previously shown to be suitable for visual cortex stimulation (28, 39, 52). The electrodes were placed on the head at least 20 min after the application of an anesthetic cream. One tACS 5x5cm rubber electrode was placed over the occipital region (3 cm above inion, Oz in the 10-20 EEG system) and one 5x7cm rubber electrode over the vertex (Cz in the 10-20 EEG system). Electroconductive gel was applied to the contact side of the rubber electrodes (NeuroConn GmbH, Ilmenau, Germany) to reduce skin impedance. The impedance between the electrodes was monitored and kept below 15 kΩ. We used electric field modelling to verify that our electrodes target V1. Simulations were run in SimNIBS 2.1 (53) using the average MNI brain template (see **Figure 3B**). Note, that the software enables finite-element modelling of electric field distribution of direct current stimulation without taking into account the temporal characteristics of the alternating current.

Since we used a very brief stimulation time (2 s only), fade in/out periods were not possible (54). Accordingly, some participants were able to distinguish the stimulation conditions (see *Results*). We accounted for this possible bias using a control measure and analysis of the potential transcutaneous sensations. In each session, before the start of the main experiment, participants were familiarized with tACS and we assessed the detectability of potential cutaneous sensations (**Figure 3D**). The detection task consisted of 20 trials. Participants received either 2s tACS (0.75, 1, and 1.5mA $tACS_{triangle}$ in experiment 1 or $tACS_{sine}$ in experiment 2) or no tACS, to test whether they can distinguish between stimulation vs no stimulation. The task after each trial was to indicate on a keyboard whether they felt a sensation underneath the tACS electrodes. In experiment 2 an additional control measurement was added to assess the potential phosphene induction by the tACS waveform. tACS (in lower frequency range) was previously suggested to induce visual phosphenes (55, 56). The protocol was the same with the only difference that this time after each trial participants indicated on a keyboard whether they perceived any visual sensations while looking on the black computer screen. The determined detection accuracy (hit rates, HR, defined as the proportion of trials in which a stimulation is present and the participant correctly responds to it) of the cutaneous sensation (experiment 1 and 2) and phosphenes (experiment 2) induced by tACS served as a control to estimate whether any unspecific effects of the stimulation might have confounded the experimental outcomes (54). In the control analysis we used the HRs for detection tACS stimulation conditions (separately for $tACS_{triangle}$ and $tACS_{sine}$) as covariates (see *Statistical Analysis*).

## 2.3 Statistical analysis

Statistical analyses were performed in IBM SPSS Statistics version 26.0 (IBM Corp.) and JASP (0.16.3) unless otherwise stated. All data was tested for normal distribution using Shapiro-Wilks test of normality. Partial eta-squared (small $\eta_p^2$ = 0.01, medium $\eta_p^2$ = 0.06, large $\eta_p^2$ = 0.14; (57)) or Cohen's d (small d=0.20–0.49, medium d=0.50–0.80, large d > 0.80; (58)) values are reported as a measure of effect-sizes. Standard statistics for simple effects were complemented with their Bayesian equivalents using the Bayes factor ($BF_{01}$) with $BF_{01}$ > 1 indicating evidence in favor of the null hypothesis over the alternative hypothesis. $BF_{01}$ were primarily provided to statistically confirm the lack of an effect throughout the analyses. Variance is reported as SD in the main text and as SE in the figures. Statistical analysis of $tACS_{triangle}$ and $tACS_{sine}$ effects was analogous to the one performed to test hf-tRNS effects (39).

## 2.3.1 Analysis of VCT modulation in $tACS_{triangle}$ and $tACS_{sine}$ experiments

First, we tested whether baseline VCT in the no tACS condition differed across the two experimental sessions using a Bayesian independent samples t-test (average baseline VCT in blocks 1-2 in experiments 1-2) using the $BF_{01}$.

For all repeated measures analysis of variance (rmANOVA) models, sphericity was assessed with Mauchly's sphericity test. The threshold for statistical significance was set at $\alpha = 0.05$. Bonferroni correction for multiple comparisons was applied where appropriate (i.e., post hoc tests; preplanned comparisons of stimulation 0.75mA, 1mA and 1.5mA vs no tACS baseline).

To test the influence of $tACS_{triangle}$ on contrast sensitivity, VCT data collected in experiment 1 ($tACS_{triangle}$) were analyzed with a rmANOVA with the factors of *tACS_{triangle}* (no, 0.75mA, 1mA, and 1.5mA $tACS_{triangle}$) and *block* (1st, 2nd). For each individual and each block, we determined the maximal behavioral improvement, i.e., lowest VCT measured when $tACS_{triangle}$ was applied, and the associated "optimal" individual $tACS_{triangle}$ intensity (ind-$tACS_{triangle}$). Note, that the ind-$tACS_{triangle}$ was always selected from active stimulation conditions (i.e., even if participants performed better in the no tACS baseline, the ind-tACS intensity was defined based on the lowest VCT during stimulation). The maximal behavioral improvements in the 1st and the 2nd block were compared using a t-test (2-tailed) for dependent measurements. Importantly, we determined ind-$tACS_{triangle}$ in the 1st block, and then used the VCT data of the separate 2nd block to test whether the associated VCT is lower compared to the no tACS condition using t-tests for dependent measures. Since we had the directional hypothesis that VCT is lower for the ind-$tACS_{triangle}$ intensity compared to no tACS this test was 1-tailed. Determining ind-$tACS_{triangle}$ and testing its effect on VCT in two separate datasets is important to not overestimate the effect of $tACS_{triangle}$ on visual detection behavior.

Similarly, VCT data collected in experiment 2 ($tACS_{sine}$) was analyzed with a rmANOVA with the factor of *tACS_{sine}* (no, 0.75mA, 1mA, and 1.5mA $tACS_{sine}$) and the factor *block* (1st, 2nd).

Again, for each individual and each block, we determined the maximal behavioral improvement and the associated ind-tACS$_{sine}$. We compared results obtained in the first and second block using the same statistical tests as for the experiment 1. The maximal behavioral improvements were compared using a t-test (2-tailed) for dependent measurements. We examined whether the ind-tACS$_{sine}$ determined based on the best behavioral performance in 1$^{st}$ block, caused VCT to be lower compared to the no tACS condition when retested on the independent dataset (2$^{nd}$ block) using t-tests (1-tailed) for dependent measures.

In both experiments to assess a general modulation of VCT induced by tACS we calculated the mean change in VCT in all active tACS conditions from 1$^{st}$ and 2$^{nd}$ blocks normalized to baseline no tACS condition (tACS-induced modulation).

To control for any potential unspecific effects of tACS we repeated the main analyses of VCT (i.e., rmANOVA) with adding HRs of cutaneous sensation for all current levels (experiment 1, tACS$_{triangle}$ and 2, tACS$_{sine}$) and phosphene detection (experiment 2, tACS$_{sine}$) as covariate. We also tested correlations between the average HR of cutaneous sensation (experiment 1 and 2) and phosphene (experiment 2) detection and average tACS-induced modulation using a Pearson correlation coefficient.

### 2.3.2 Comparison of stimulation-induced VCT modulation in tACS$_{triangle}$, tACS$_{sine}$, and hf-tRNS experiments

We compared the effects of deterministic transcranial electrical stimulation (tES, i.e., tACS$_{triangle}$ and tACS$_{sine}$) and stochastic tES (i.e., hf-tRNS) on VCT. The data demonstrating the effect of hf-tRNS on VCT were taken from a previous study investigating the effects of hf-tRNS using the same behavioral paradigm (39).

First, we tested whether baseline VCT in the no tES (no tACS, no hf-tRNS) conditions differed across the three experiments using a Bayesian independent samples t-test (average baseline VCT in blocks 1-2 in $tACS_{triangle}$, $tACS_{sine}$ and hf-tRNS) using the $BF_{01}$.

Next, we tested whether a general tES-induced modulation of VCT (mean of all active stimulation conditions from two blocks normalized to baseline no stimulation condition) differed across the three experiments using a Bayesian ANOVA (tES-induced modulation in $tACS_{triangle}$, $tACS_{sine}$ and hf-tRNS experiments) using the $BF_{01}$.

Finally, we depicted tES-induced modulation of VCT as paired Cohen's d bootstrapped sampling distributions employing an online tool (https://www.estimationstats.com; (59)). For each pair of control no tES (i.e., no tACS in $tACS_{triangle}$, $tACS_{sine}$ and no hf-tRNS) and tES conditions ($tACS_{triangle}$, $tACS_{sine}$, hf-tRNS) a two-sided permutation t-tests were conducted. 5000 bootstrap samples were taken. The confidence interval was bias-corrected and accelerated. The reported P values are the likelihoods of observing the effect sizes, if the null hypothesis of zero difference is true. For each permutation P value, 5000 reshuffles of the control and test labels were performed.

## 3. Results

We first tested whether VCT measured during the no tACS conditions differed between the experiments (i.e., average baseline VCT in $tACS_{triangle}$ and $tACS_{sine}$ experiments, see **Figure 4**). Bayesian independent samples t-test revealed that the baseline VCT measured in the no tACS condition did not differ between experiments ($BF_{01}$ = 3.439, i.e., moderate evidence for the $H_0$).

### FIGURE 4 ###

### 3.1 $tACS_{triangle}$ over V1 modulates visual contrast threshold

In the first experiment, we investigated whether $tACS_{triangle}$ modulates the visual contrast detection when applied to V1. We measured VCT during $tACS_{triangle}$ at intensities of 0.75, 1, to 1.5mA peak-to-baseline versus no tACS control condition. We found a general decrease in VCT ($F_{(3, 81)}$ = 3.41, p = 0.021, $\eta_p^2$ = 0.11, $BF_{01}$ = 0.498) reflecting improved contrast sensitivity during $tACS_{triangle}$ (**Figure 5A**). Post hoc comparisons revealed that 0.75mA and 1mA stimulation were most effective in boosting contrast processing at a group level, which differed significantly from the no tACS control condition (p = 0.033, mean difference, MD = -6.3 $\pm$ 11.62% and p = 0.024, MD = -6.33 $\pm$ 10.45%, respectively). Neither the main effect of *block* ($F_{(1, 27)}$ = 2.43, p = 0.13, $BF_{01}$ = 1.429) nor $tACS_{triangle}$*block* interaction ($F_{(3, 81)}$ = 1.6, p = 0.195) reached significance.

When comparing $tACS_{triangle}$-induced effects between the 1st and 2nd block we found that the maximal behavioral improvement (i.e., maximal $tACS_{triangle}$-induced lowering of the VCT relative to the no tACS condition) were not significantly different between the 1st (MD = -14.64 $\pm$ 12.6%, VCT decrease in 25 out of 28 individuals) and the 2nd block (MD = -15.75 $\pm$ 15.73%, VCT decrease in 24 out of 28 individuals; $t_{(27)}$ = 0.604, p = 0.551, $BF_{01}$ = 4.219), additionally showing that no time effects arose from the first to the second block of measurement.

Next, we defined the optimal $ind\text{-}tACS_{triangle}$ for each participant and examined whether its effects can be reproduced. We observed that the $ind\text{-}tACS_{triangle}$ determined in 1st block (**Figure 5B**) caused decrease in VCT compared to the no tACS condition when retested within the same experimental session ($t_{(27)}$ = 1.84, p = 0.039, $BF_{01}$ = 0.463, VCT decrease in 18 out of 28 individuals, MD = -5.26 $\pm$ 18.23%, **Figure 5C**). Note, that the above analysis does not contain an element of intrinsic circularity because the $ind\text{-}tACS_{triangle}$ and the VCT measure were based on independent data sets.

The cutaneous sensation control experiment revealed that some of our participants could detect $tACS_{triangle}$ conditions (HR at 0.75mA = 12.5 $\pm$ 25%, 1mA = 18.75 $\pm$ 27.74%, 1.5mA =

$41.07 \pm 43.68\%$, mean HR = $24.11 \pm 27.34\%$). We reanalyzed our main outcome parameter by adding sensation detection HRs for each current level as covariates (HRs were z-scored because of non-normal distribution). The main effect of $tACS_{triangle}$ remained significant ($F_{(3, 72)}$ = 3.36, p = 0.023, $\eta_p^2$ = 0.12). Moreover, the mean HR of cutaneous sensation detection did not correlate with the average $tACS_{triangle}$-induced VCT modulation (r = 0.181, p = 0.357, $BF_{01}$ = 2.835), making it unlikely that transcutaneous sensation was the main driver of our results.

### FIGURE 5 ###

**3.2 $tACS_{sine}$ over V1 modulates visual contrast threshold**

In the second experiment, we explored the effects of $tACS_{sine}$ applied over V1 on visual contrast detection. VCT was measured during $tACS_{sine}$ at intensities of 0.75, 1, to 1.5mA peak-to-baseline versus no tACS control condition. We observed a general decrease in VCT with increasing $tACS_{sine}$ intensity ($F_{(3, 81)}$ = 4.78, p = 0.004, $\eta_p^2$ = 0.15, $BF_{01}$ = 0.111) reflecting improved contrast sensitivity during $tACS_{sine}$. Post hoc comparisons revealed that the 1mA and 1.5mA stimulation were most effective in enhancing contrast processing, which differed significantly from the no tACS control condition (p = 0.042, MD = $-8.04 \pm 13.82\%$ and p = 0.008, MD = $-6.52 \pm 12.66\%$, respectively, **Figure 6A**). There was no main effect of *block* ($F_{(1, 27)}$ = 0.02, p = 0.878, $BF_{01}$ = 3.619) or $tACS_{sine}$*block* interaction ($F_{(3, 81)}$ = 0.5, p = 0.684).

When comparing $tACS_{sine}$-induced effects between the 1st and 2nd block we found that the maximal behavioral improvement, defined as maximal $tACS_{sine}$ induced lowering of the VCT were not different between the 1st (MD = $-17.78 \pm 15.82\%$, VCT decrease in 25 out of 28 individuals) and the 2nd block (MD = $-18.37 \pm 16.67\%$, VCT decrease in 22 out of 28 individuals; $t_{(27)}$ = 0.95, p = 0.353, $BF_{01}$ = 3.320).

We determined the optimal ind-tACS$_{sine}$ and tested whether its effects can be reproduced. Similar to ind-tACS$_{triangle}$ in experiment 1, the optimal ind-tACS$_{sine}$ determined in 1$^{st}$ block (**Figure 6B**) significantly lowered the VCT compared to the no tACS condition when retested on the independent VCT data set of the 2$^{nd}$ block (t$_{(27)}$ = 2.59, p = 0.008, BF$_{01}$ = 0.157, VCT decrease in 18 out of 28 individuals, MD = -7.85 $\pm$ 21.84%, **Figure 6C**).

Similarly to experiment 1, we assessed the HR of cutaneous sensation detection (HR at 0.75mA = 16.07 $\pm$ 27.4%, 1mA = 21.43 $\pm$ 30.21%, 1.5mA = 50.89 $\pm$ 43.29%, mean HR = 29.46 $\pm$ 27.36%). We reanalyzed our main outcome parameter by adding mean cutaneous sensation detection HRs as a covariate (HRs were z-scored because of non-normal distribution). The main effect of *tACS$_{sine}$* remained significant (F$_{(3, 72)}$ = 4.67, p = 0.005, $\eta_p^2$ = 0.16). The mean HR of cutaneous sensation did not correlate with the average tACS$_{sine}$-induced VCT modulation (r = -0.12, p = 0.542, BF$_{01}$ = 3.569). In this experiment we additionally tested phosphenes detection (HR$_{phos}$ at 0.75mA = 3.57 $\pm$ 8.91%, 1mA = 5.36 $\pm$ 12.47%, 1.5mA = 6.25 $\pm$ 16.14%, mean HR = 5.06 $\pm$ 10.48%). After adding HR$_{phos}$ as covariate (z-scored HR$_{phos}$), the main effect of *tACS$_{sine}$* remained significant (F$_{(3, 72)}$ = 4.82, p = 0.004, $\eta_p^2$ = 0.17). Accordingly, the mean HR of phosphene detection did not correlate with the average tACS$_{sine}$-induced VCT modulation (r = -0.14, p = 0.493, BF$_{01}$ = 3.405).

### FIGURE 6 ###

### 3.3 Comparison of tACS$_{triangle}$, tACS$_{sine}$, and hf-tRNS-induced modulation

First, we tested whether baseline VCT measured during the no tES conditions differed between the experiments (i.e., average baseline VCT in tACS$_{triangle}$, tACS$_{sine}$, and hf-tRNS experiments, see **Figure 4**). Bayesian independent samples t-test revealed that the baseline VCT measured in the no tES condition did not differ between experiments (BF$_{01}$ = 4.869, i.e., moderate evidence for the H$_0$).

Next, we compared tES-induced modulation effects between experiments (tACS$_{triangle}$, tACS$_{sine}$ and hf-tRNS experiments, see **Figure 7A**). A Bayesian ANOVA revealed that the general tES-induced modulation did not differ between experiments (BF$_{01}$ = 8.956, i.e., moderate evidence for the H$_0$), suggesting that all three stimulation types were equally effective in lowering VCT.

Finally, we assessed the strength of the tES-induced effects on VCT across tACS$_{triangle}$, tACS$_{sine}$ and hf-tRNS experiments defined as paired Cohen's d bootstrapped sampling distributions (see **Figure 7B**). We found comparable (small) effects of significant differences between no tES baseline VCT and averaged VCT in active tES conditions in all experiments using the two-sided permutation t-test [in tACS$_{triangle}$ d = -0.17 (95.0%CI -0.284; -0.0698) p = 0.0034; in tACS$_{sine}$ d = -0.242 (95.0%CI -0.444; -0.103), p = 0.0016; in hf-tRNS d = -0.249 (95.0%CI -0.433; -0.088) p = 0.0092]. The effect sizes and CIs are reported above as: effect size (CI width lower bound; upper bound).

### FIGURE 7 ###

## 4. Discussion

Theoretical modelling shows that adding a deterministic, high-frequency sinusoidal signal instead of stochastic noise could lead to signal enhancement due to resonance, according to the DAR mechanism. Our experimental proof-of-concept study revealed, that stimulation of V1 with a deterministic tACS signal instead of stochastic noise leads to signal enhancement in visual processing. We measured visual contrast sensitivity during tACS$_{triangle}$ and tACS$_{sine}$. On the group level, we found consistent tACS$_{triangle}$- and tACS$_{sine}$-induced decrease in VCT, reflecting enhancement in visual contrast processing during V1 stimulation (**Figure 5A**, **Figure 6A**). The online modulation effects of individually optimized tACS$_{triangle}$ and tACS$_{sine}$ intensities (**Figure 5B**, **Figure 6B**) were replicated on the independent VCT data (**Figure**

**5C**, **Figure 6C**). Finally, we demonstrated that the effects of deterministic stimulation on VCT are comparable to stochastic stimulation of V1 with hf-tRNS (**Figure 7AB**).

## 4.1 tACS with triangle and sine waveform improve visual sensitivity

Our findings provide the first proof of concept that the deterministic $tACS_{triangle}$ and $tACS_{sine}$ delivered to V1 can modulate visual contrast sensitivity. Across two experiments we showed that the modulatory effects of tACS on visual sensitivity are not waveform specific, as both $tACS_{triangle}$ and $tACS_{sine}$ induced significant decrease in VCT (**Figure 5A**, **Figure 6A**).

One of the main characteristics of SR-like effects is the optimal intensity of noise, which is required in order to yield the improved performance (5, 15). Here, we did not observe an excessive level of tACS that would be detrimental for visual processing (**Figure 5A**, **Figure 6A**). This is consistent with our predictions that, in line with DAR (see mathematical predictions in 1.2.1), adding high frequency deterministic signal should result in a noise-free output where the detection processing is not disturbed by random stimulation effects.

Similar to other studies investigating resonance-like effects (28, 39), our results have revealed large variability among participants in terms of the optimal intensity resulting in the strongest modulation of visual contrast sensitivity (**Figure 5B**, **Figure 6B**). However, consistent with the effects of tRNS-induced online modulation of contrast processing in V1 shown previously (28, 39), the effects of individualized tACS intensity were replicated on the independent VCT data set collected within the same experimental session (**Figure 5C**, **Figure 6C**), suggesting consistent beneficial influence of tACS on signal enhancement.

We implemented several control measures to test whether the improvement in visual processing was driven by effective stimulation of V1 rather than any unspecific effects of tACS. We applied an anesthetic cream to numb potential stimulation-induced cutaneous sensation on the scalp (50, 51). While the anesthetic cream numbs the skin and reduces the cutaneous sensations resulting from tACS, it does not eliminate them completely in all individuals. The control cutaneous sensation detection assessment in the current study showed that some participants could accurately detect tACS, and that the mean detection rate was rather low (mean HR = 24.11 ± 27.34% in tACS$_{triangle}$ and mean HR = 29.46 ± 27.36% in tACS$_{sine}$). Cutaneous sensation and phosphenes detection (also very low, mean HR$_{phos}$ = 5.06 ± 10.48%) did not correlate with the average tACS-induced VCT modulation neither in tACS$_{triangle}$, nor tACS$_{sine}$ experiment. Moreover, stimulation effects remained significant in the additional analysis using tactile or phosphene sensation detection during tACS$_{triangle}$ and tACS$_{sine}$ as covariate.

While tACS$_{sine}$ is a well-established and frequently used non-invasive brain stimulation method, high frequency tACS$_{sine}$ is less common. The effects of 80Hz tACS$_{sine}$ were sporadically tested in the past using physiological and behavioral paradigms. Ten minutes of 140Hz tACS$_{sine}$ was shown to increase primary motor cortex (M1) excitability as measured by transcranial magnetic stimulation-elicited motor evoked potentials during and for up to 1h after stimulation. Control experiments with sham and 80Hz stimulation did not show any effect, and 250Hz stimulation was less efficient, with a delayed excitability induction and reduced duration (60). The researchers postulated that the changes in corticospinal excitability result from externally applied high frequency oscillation in the ripple range (140Hz corresponding to middle, 80Hz lower and 250Hz upper border) that interfere with ongoing oscillations and neuronal activity in the brain (60). We can, however, not directly translate the effects of tACS$_{sine}$ of M1 to our stimulation of V1. Additionally, the stimulation effects observed in our study are likely reflecting acute modulation of contrast processing, as stimulation was only applied for short intervals (2 s) always interleaved with control (no tACS) condition. Thus, it is possible that even though 80Hz stimulation did not lead to long term effects in cortical excitability it can still affect cortical processes acutely.

In the visual domain, 1.5mA high-frequency tACS$_{sine}$ was applied to V1 for 15-45min in a study investigating the effect of covert spatial attention on contrast sensitivity and contrast discrimination (61). That study found that contrast discrimination thresholds decreased significantly during 60Hz tACS$_{sine}$, but not during 40 and 80Hz stimulation. This previous study used, however, different visual stimuli than that utilized here, i.e., a random dot pattern. Moreover, they used a more complicated behavioral paradigm, where contrast-discrimination thresholds were tested using two attention conditions, i.e., with or without a peripheral cue, as the study goal was to explore the influence of attentional processes on visual tasks. One tACS mechanism that has been tested in the visual domain is the reduction of adaptation (62, 63). More specifically, a seminal study has shown that 10Hz tACS reduces sensory adaptation in a visual motion perception task (62). Since sensory adaptation increases thresholds for detection, potentially reducing adaptation during tACS could result in decreased thresholds. However, the aforementioned study design differed substantially from the one presented here. Kar and Krekelberg (62) used a 40s adaptor stimulus to induce adaptation while our stimuli were presented for only 40ms (i.e., 3 magnitudes shorter). This is relevant because it was previously shown that adaptation gets stronger and lasts longer with increasing adaptation duration (64). Moreover, Kar and Krekelberg (62) used 10Hz tACS which was substantially lower than the one exploited in our experiments (i.e., 80Hz), making direct comparisons difficult. Overall, additional experiments would be required to test whether our results could be explained by a reduction of visual adaptation.

Even though the vast majority of tACS studies to date have used a sinusoidal waveform, an alternating current does not have to be sinusoidal, since it can take any arbitrary waveform such as rectangular wave (65), pulsed (66), or sawtooth (67). Dowsett and Herrmann (2016) investigated the effects of sinusoidal and sawtooth wave tACS on individual endogenous alpha-power enhancement. They observed alpha oscillations enhancement both during and after sawtooth stimulation. The effect seemed to depend on the shape of the sawtooth, as they found that positive, but not negative, ramp sawtooth significantly enhanced alpha power during stimulation relative to sham. They postulated that a sudden, instantaneous change in current might be more effective than a sinusoidal current in increasing the probability of neurons firing. In this regard, Fröhlich and McCormick (Supplementary Material in (68)) demonstrated that ramps of increasing voltage with a steeper gradient resulted in increased neural firing in vitro, relative to ramps with a low gradient but reaching the same maximum voltage. This suggests that it is not only the total amount of current but also the rate of change of current can modulate neural firing. Note, that triangle waveform has a faster rate of change of current than the sine wave.

Although we postulate that the effect of tACS on VCT in our study results from resonance-like mechanism, this is not the only potential mechanism. Importantly, the commonly accepted mechanism of action of tACS is that it entrains action potential firing, and thus neural oscillations (69). Entrainment effect anticipates a monotonic relationship between the tACS effect and intensity, where increasing stimulation intensity results in greater effects for stimulation waveforms that are tuned to the endogenous oscillation (70). The effects of tACS in regard to induced brain oscillations seem to depend on the stimulation duration (71). Although the entrainment after-effects were observed after tACS had been delivered for several minutes (71), short stimulation of 1s did not produce after-effects on amplitude or phase of the electroencephalogram (72). Moreover, in the study investigating the effects of intermittent alpha tACS of either 3 or 8 s, the after-effects were found only for the 8-s condition (73). The authors excluded entrainment as potential underlying mechanism and postulated plasticity-related changes as the responsible mechanism for the observed after-effects. Here, we used very brief stimulation of 2s tACS per trial, a duration seemingly too short to induce the entrainment effects on cortical processing.

Furthermore, it was postulated that a very small amount of applied electric field can bias spike timing or spike probability when a neuron nears the threshold of spike generation (74). Accordingly, it was shown that although entrainment effects can arise at field strengths <0.5 mV/mm, physiological effects are more pronounced for higher intensities (around 1mV/mm), according to intracranial recordings in awake nonhuman primates (75). These values are well above the simulated induced electric field in our study (around 0.2 mV/mm, see **Figure 3**). Further studies are required to fully disentangle the underlying neuronal effects of tACS driving the enhancement in visual detection. To exclude the influence of entrainment on VCT modulation a jittered tACS protocol could be employed. A paradigm using stimulation of jittered flickering light, where instead of a rhythmic flicker, inter stimulus intervals of the square wave were jittered with a maximum of ± 60%, was shown to fail in inducing rhythmic brain response (76). If a jittered tACS of V1 would still influence contrast sensitivity we could assume the non-entrainment origin of the effect.

## 4.2 Comparison of tACS$_{triangle}$, tACS$_{sine}$, and hf-tRNS

In the phenomenon of SR, random noise added to a non-linear system can increase its responsiveness towards weak subthreshold stimuli. One aim in the present study was to explore whether a deterministic and periodic signal can substitute stochastic noise and still lead to response enhancement in a threshold-based stochastic resonator. The DAR characteristics of high-frequency deterministic signal might offer a noise-free output, thus additionally increasing SNR. We proposed the following testable hypotheses (i) tACS$_{triangle}$ will have a larger resonance-like effect compared to hf-tRNS, (ii) tACS$_{sine}$ will have less effect than tACS$_{triangle}$, due to the loss of waveform linearity. We found enhancement effects of both tACS$_{triangle}$ vs tACS$_{sine}$ (**Figure 5A**, **Figure 6A**), however to test whether these effects are indeed superior to stochastic stimulation, we directly compared the VCT modulation induced by tACS$_{triangle}$, tACS$_{sine}$ and hf-tRNS (**Figure 7**). The baseline contrast sensitivity between the compared experiments was not different (**Figure 4**). Counter to our hypothesis, the noise-free tACS did not result in stronger contrast sensitivity enhancement, as average VCT modulation did not differ between the three stimulation conditions, as confirmed by Bayesian analysis (**Figure 7A**). Accordingly, the effects sizes of all three stimulation types were comparable (**Figure 7B**). Therefore, we showed that both deterministic and stochastic high-frequency stimulations were equally effective in inducing resonance-like effects.

In real life (in comparison to mathematical simulations) neural processing is intrinsically noisy. How this intrinsic noise interacts with the applied noise/SR signal will have implications for the validity of our mathematical model. The task we used is a 4AFC discrimination task, which means that when tRNS is added to the neurons that there must be a distinction between 3 noisy locations and 1 signal and noise location. How added noise influences this comparison remains unresolved and an area for further investigation. To investigate this, a two-sided tRNS experiment could be run where noise is added to the left and/or right V1 (or S1) to explore the influence of more noise added to the system where the signal is not present versus when it is present. This might give some information on how the brain interacts with signal, added noise and intrinsic noise compared to intrinsic noise only on a discrimination task.

### 4.3 Conclusions

The present study provides the first evidence for resonance-like neural signal enhancement without adding a stochastic noise component. We showed that 'deterministic' 80Hz-tACS and 'stochastic' hf-tRNS are equally effective in enhancing visual contrast detection. In the range of commonly used intensities of tES to induce SR, tACS did not result in detrimental effects related to excessive interference signal, thus providing increased SNR in all tested intensities, according to DAR predictions. These findings shed a new light on the effects induced by both 80 Hz tACS and hf-tRNS, and their underlying mechanisms.

The optical excitations in this work were square wave signals (i.e., the visual stimulus was switched on for 40 ms). Open theoretical and related experimental questions are: What is the role of the finite time it takes for the square wave signal to reach its amplitude? What if the signal has a different periodicity? Possible interrelations between the pulses and the inter-pulse intervals?

# GLOSSARY

**Stochastic Resonance (SR)** – certain nonlinear systems show improved signal transfer in the presence of high-frequency additive noise. It is an amplitude resonance because there is an optimal noise (root mean square) amplitude for the best transfer.

**Threshold Elements (TE)** – a device with threshold-based nonlinearity.

**Level Crossing Detector (LCD)** – a threshold element that produces a short uniform spike signal at its output whenever the input signal crosses the threshold level. There are variations depending on what type of crossing (up, down, or both) triggers a spike.

**Comparator** – a threshold element that produces zero output value whenever the input signal is below the threshold and a non-zero $U_H$ value otherwise.

**Signal strength (SS)** - the mean-square amplitude of the signal.

**Signal-to-noise-ratio (SNR)** - the ratio of the mean-square amplitudes of signal and noise.

**Information entropy** - the maximum of the useful information that an unknown message with a given size can contain.

**Shannon information channel capacity** - the maximum bit rate that a (typically noisy) information channel can effectively transfer.

**Deterministic Amplitude Resonance (DAR)** - a device that, similarly to SR, show improved signal transfer in the presence of high-frequency, additive, deterministic, carrier-wave. It is an amplitude resonance because there is optimal carrier-wave amplitude for the best transfer.

**Periodic carrier-waves** - carrier waves that are periodic time functions.

**Triangle wave** - periodic carrier-wave with straight lines of the rising and the falling sections.

**Sine wave** - sinusoidal carrier-wave

**Transcranial Electric Stimulation (tES)** – noninvasive brain stimulation technique, which applies weak, painless electrical currents to the scalp (current intensities of ~1–2 mA), to modulate brain function (77–80).

**Transcranial Alternating Current Stimulation (tACS)** - a subtype of tES characterized by biphasic, alternating electric currents applied (69, 81).

**Transcranial Random Noise Stimulation (tRNS)** - a subtype of tES whereby currents are randomly drawn from a predefined range of intensities and frequencies (26, 82, 83).

**Visual contrast detection threshold (VCT)** – criterion reflecting the level of task performance accuracy. Here, the detection threshold corresponds to the contrast intensity of presented visual stimuli that was accurately detected with 50% accuracy (see Figure 3).

**QUEST staircase procedure** - a method used in psychophysical research to estimate threshold of a psychometric function (47). In this maximum-likelihood adaptive procedure, information from all trials in an experiment are considered to determine a threshold (47, 48). Here, QUEST method was used to estimate the visual contrast detection threshold for each participant.

**Four-alternative forced choice (4-AFC) visual task** - design of a discrimination task in psychophysical experiments, where participant is forced to choose one out of four possible responses. In contrast to methods requiring a 'yes/no' response, forced-choice methods characterize with higher accuracy of the measured psychophysical property (45). Here, the weak visual stimulus was presented with different intensities in one of 4 quadrants on the screen and participants were asked to select in which one it appeared in each trial. Based on the accuracy of those responses we estimated their contrast detection threshold using QUEST procedure.

**Figures Captions**

**Figure 1** Deterministic transfer of sub-threshold binary signal through simple threshold-based stochastic resonators with a Threshold Element (TE: either a Level Crossing Detector (LCD) or a Comparator) and an additive triangle wave at the input. Note: the classical threshold-based stochastic resonators contain the same hardware elements except the triangle wave that is substituted by a Gaussian random noise. The role of the Low-pass Filter is to reduce the amount of irrelevant high-frequency products created by the carrier wave. If those irrelevant high-frequency products are not disturbing, the Low-pass Filter can be omitted. Upper part: the sub-threshold binary signal is unable to trigger the TE thus the output signal is steadily zero. Lower part: an additive, triangle wave (carrier-wave) assists the signal to reach the threshold thus it carries the binary signal over the TE. The Low-pass filter takes a short time average in order to smooth out the high-frequency components. For high-fidelity transfer, to avoid problems caused by delays or phase shifts, the frequency of the carrier-wave must be much greater than that of the binary signal. In the old stochastic resonance schemes, the carrier-wave was a noise that caused a non-deterministic component (noise) and finite SNR at the output. The new system is purely deterministic, and its SNR is infinite. Moreover, if the signal is "analog" (continuum amplitude values), the triangle wave with comparator as TE guarantees a linear transfer of the signal provided the threshold level is between the minimum and the maximum of the sum of the signal and the carrier-wave, see in (ii) below. $U_{th}$ = threshold, $U_s$ = signal, $U_t$ = noise, $U_{lcd}$ = LCD output signal, $U_c$ = comparator output signal, $U_{LPF}$ = signal after low-pass filtering.

**Figure 2** The triangle wave vs. the threshold ($U_{th}$).

**Figure 3** Experimental design. **A.** Example trial of 4-alternative forced choice task measuring visual contrast detection threshold (VCT). tACS was delivered for 2 s around the Gabor patch presentation. **B.** tACS electrodes montage targeting V1 and simulation of the induced electric field in the brain. **C.** Example of dose-response psychometric curves and the VCT for the 50% detection accuracy level. We hypothesize that the VCT will be lower (indicating better contrast detection performance of the participant) in one of the tACS conditions (violet) than in the no tACS control condition (blue). **D.** The order of measurements within each experiment. Each experimental session consisted of application of an anesthetic cream, followed by task training, familiarization protocol, and two independent VCT assessments in 4 interleaved tACS conditions (as specified in A).

**Figure 4** Average baseline VCT measured in the no tES conditions in $tACS_{triangle}$ N=28 (16 females, 12 males), $tACS_{sine}$ N=28 (20 females, 8 males), hf-tRNS experiments N=24 (16 females, 8 males). VCT was assessed for stimuli presented with contrast intensity ranging from 0 to 1. Blue lines indicate mean, gray dots indicate single subject data. $BF_{01}$ = 4.869, i.e., moderate evidence for the $H_0$.

**Figure 5** The effect of $tACS_{triangle}$ on VCT measured in experiment 1. VCT was assessed for stimuli presented with contrast intensity ranging from 0 to 1. **A.** Effect of $tACS_{triangle}$ on VCT on a group level measured across $1^{st}$ and $2^{nd}$ blocks. Decrease in VCT reflects improvement of visual contrast sensitivity. VCT in $tACS_{triangle}$ conditions normalized to the no stimulation baseline. All data mean ± SE; *p < 0.05, rmANOVA **B.** Individually defined optimal $tACS_{triangle}$ based on behavioral performance during the $1^{st}$ block. **C.** Detection improvement effects of individualized $tACS_{triangle}$ (selected based on block 1) measured on the independent VCT data of block 2. Gray dots indicate single subject data; *p < 0.05, t-test for dependent measures. N=28 (16 females, 12 males).

**Figure 6** The effect of $tACS_{sine}$ on VCT measured in experiment 2. VCT was assessed for stimuli presented with contrast intensity ranging from 0 to 1. **A.** Effect of $tACS_{sine}$ on VCT on a group level measured across $1^{st}$ and $2^{nd}$ blocks. Decrease in VCT reflects improvement of visual contrast sensitivity. VCT in $tACS_{sine}$ conditions normalized to the no stimulation baseline. All data mean ± SE; *p < 0.05, **p < 0.01, rmANOVA. **B.** Individually defined optimal $tACS_{sine}$ based on behavioral performance during the $1^{st}$ block. **C.** Detection improvement effects of individualized $tACS_{sine}$ (selected based on block 1) measured on the independent VCT data of block 2. Gray dots indicate single subject data; **p < 0.01, t-test for dependent measures. N=28 (20 females, 8 males).

**Figure 7** Comparison of tACS$_{triangle}$ (N=28; 16 females, 12 males), tACS$_{sine}$ (N=28; 20 females, 8 males), and hf-tRNS-induced modulation (N=24; 16 females, 8 males). **A.** VCT modulation induced by tACS$_{triangle}$, tACS$_{sine}$, hf-tRNS. The general modulation of VCT induced by tES was calculated as mean of all active tES conditions from $1^{st}$ and $2^{nd}$ blocks normalized to baseline no tES condition in each experiment. Decrease in VCT reflects improvement of visual contrast sensitivity. **p < 0.01, two-sided permutation t-test. **B.** The paired Cohen's d for 3 comparisons shown in the Cumming estimation plot. Each paired mean difference is plotted as a bootstrap sampling distribution. Mean differences are depicted as dots, 95% confidence intervals are indicated by the ends of the vertical error bars.